\documentclass[12pt]{revtex4}
\usepackage{epsfig,url}
\usepackage{lineno}
\newcommand\oo{\omega_{12}}
\newcommand\ooe{\omega_{E}}
\newcommand\Do{\Delta\omega}
\newcommand\ff{f_{12}}
\newcommand\Df{\Delta f}
\newcommand\dt{\Delta t}
\begin{document}
%\linenumbers
\title{Simulations of real-time system identification for superconducting
  cavities with a recursive least-squares algorithm}
\author{Volker Ziemann}
\affiliation{Uppsala University, 75120 Uppsala, Sweden}
\date{\today}
%%%%%%%%%%%%%%%%%%%%%%%%%%%%%%%%%%%%%%%%%%%%%%%%%%%%%%%%%%%%%%%%%%%%%%%
\begin{abstract}\noindent
  We explore the performance of a recursive least-squares algorithm to determine the
  bandwidth $\omega_{12}$ and the detuning $\Delta\omega$ of a superconducting cavity.
  We base the simulations on parameters of the ESS double-spoke cavities. Expressions
  for the signal-to-noise ratio of derived parameters are given to explore the
  applicability of the algorithm to other configurations. 
\end{abstract}
\maketitle
%
%%%%%%%%%%%%%%%%%%%%%%%%%%%%%%%%%%%%%%%%%%%%%%%%%%%%%%%%%%%%%%%%%%%%%%%
%
\section{Introduction}
\label{sec:intro}
Superconducting accelerating cavities are used to accelerate protons~\cite{SNS,ESS},
electrons~\cite{CEBAF,XFEL,CBETA}, and heavy ions~\cite{FRIB,SPIRAL2,HELIAC}, both
with pulsed~\cite{SNS,XFEL} and with continuous beams~\cite{CEBAF,JLABFEL}. Owing
to the low losses, the cavities have a very narrow bandwidth on the order of Hz for
bare cavities and a few 100\,Hz for cavities equipped with high-power couplers.
In order to efficiently cool these cavities with liquid helium they are
made of rather thin material, which makes them easily deformable and this changes
their resonance frequency, often by an amount comparable to their bandwidth.
In pulsed operation, the dominant deformation comes from the electro-magnetic
pressure of the field inside the cavity, the Lorentz-force detuning~\cite{LFD,ACE3},
while cavities operating continuously are perturbed by so-called microphonics~\cite{ANA1,NEUMANN},
caused by pressure variations of the liquid helium bath or mechanical perturbations,
for example, by reciprocating pumps or by malfunctioning equipment. As a consequence
of these perturbations, the cavities are detuned and force the power generators to
increase their output to maintain fields necessary for stable operation of the beams.
This reduces the efficiency of the system and requires an, often substantial, overhead
of the power generation, forcing it to operate at a less than optimal working point.
To avoid this sub-optimal mode of operation and  to compensate the detuning, many
accelerators employ active tuning systems that use stepper motors and
piezo-actuators~\cite{TUNER} to squeeze the cavities back in tune, which requires
diagnostic systems to measure the detuning.
%One way to determine the detuning
%employs a so-called self-excited loop, where a signal form
These measurements are usually
based on comparing the phase of the signal that excites the cavity, measured with
a directional coupler just upstream of the input coupler, to the phase of the
field inside the cavity, measured by a field probe or antenna inside the cavity.
Both analog~\cite{ANA1,ANA2} and digital~\cite{SCHILCHER,PLAWSKI} signal processing
systems are used; often as part of the low-level radio-frequency (LLRF) feedback
system that stabilizes the fields in the cavity. Even more elaborate systems,
based on various system identification algorithms, are used or
planned~\cite{RYBA,CZARSKI,BELL,ECHE}. All these algorithms normally rely on
low-pass filtering the often noisy signals from the directional couplers and
antennas in order to provide a reliable estimate of the cavity detuning and
the bandwidth.
\par
In this report, we focus on a complementary algorithm that continuously improves
the estimated fit parameters by increasing the size of a system of equations.
Instead of solving this rapidly growing system directly, we employ a recursive
least-squares (RLS) algorithm~\cite{AW,ZZ1}, which only requires moderate numerical expenditure
in each time step. Remarkably, asymptotically the difference between the continuously
improving estimates of the fit parameters and the ``true'' values---the so-called
estimation error---approaches zero~\cite{LAIWEI} albeit at the expense of a limited
ability to resolve changing parameters. We therefore introduce a finite memory
when solving the system, which downgrades old measurements in favor of new ones.
This allows us to handle even changing parameters at the expense
of an increased noise level of the fit parameters.
\par
In the following sections, we first introduce the model of the cavity and transform
the continuous-time model to discrete time. In Section~\ref{sec:sysid} we develop
the RLS algorithm to identify the cavity parameters. In Section~\ref{sec:sim} we
explore the capabilities of the algorithm in simulations before calculating the
signal-to-noise ratio in Section~\ref{sec:SN} and the conclusions.
\section{Model}
\label{sec:model}
Accelerating cavities can be described by an equivalent circuit composed of a
resistor $R$, an inductance $L$, and a capacitor $C$, all connected in parallel.
This circuit is then excited by a current $I$ and responds by a building up
a voltage $V=V_r+iV_i$ across the components. This voltage is decomposed into
real (in-phase, I) and imaginary (out-of-phase, Q) components. After averaging
over the fast oscillations the evolution of the real and imaginary parts of the
voltage envelope is given by the following state-space representation~\cite{SCHILCHER}
\begin{equation}\label{eq:sss}
  \left(\begin{array}{c} \frac{dV_r}{dt}\\ \frac{dV_i}{dt} \end{array}\right)
  =\left(\begin{array}{cc} -\oo & -\Do \\ \Do& -\oo \end{array}\right)
  \left(\begin{array}{c} V_r \\ V_i \end{array}\right)
  +\left(\begin{array}{cc} \oo R & 0 \\ 0& \oo R \end{array}\right) 
  \left(\begin{array}{c} I_r \\ I_i \end{array}\right)
\end{equation}
of the system that describes the dynamics of the cavity voltage powered by a
generator that provides the currents. The directional couplers used to measure
the input signal, however, measure the forward component of the current
$\vec I^+$ rather than the total current $\vec I=\vec I^+ + \vec I^-$. Close
to resonance, it is  straightforward to show that the measured forward current
$\vec I^+$, which proportional to the signal from the directional coupler, is
related to the total current $\vec I$ by
\begin{equation}\label{eq:curr}
  \vec I=\frac{2\beta}{1+\beta} \vec I^+ = \frac{2Q_L}{Q_E} \vec I^+ 
\end{equation}
with the coupling factor $\beta=Q_0/Q_E$ given by the ratio of the intrinsic
quality factor of the cavity $Q_0$ and the external quality factor $Q_E$. Moreover,
$1/Q_L=1/Q_0+1/Q_E=(1+\beta)/Q_0$ defines the loaded quality factor $Q_L$. Replacing the
currents on the right-hand side of Equation~\ref{eq:sss} with the help of
Equation~\ref{eq:curr} then leads to
\begin{equation}\label{eq:ss}
  \left(\begin{array}{c} \frac{dV_r}{dt}\\ \frac{dV_i}{dt} \end{array}\right)
  =\left(\begin{array}{cc} -\oo & -\Do \\ \Do& -\oo \end{array}\right)
  \left(\begin{array}{c} V_r \\ V_i \end{array}\right)
  +\left(\begin{array}{cc} \ooe R & 0 \\ 0& \ooe R \end{array}\right) 
  \left(\begin{array}{c} I^+_r \\ I^+_i \end{array}\right)
\end{equation}
with $\ooe=\hat\omega/Q_E$. We also introduce $\oo=\hat\omega/2Q_L$ and the cavity
resonance frequency $\hat\omega$. Furthermore, we assume that magnitude and
phase of all currents and voltages can be reliably measured after the hardware
(antennas, cables, and amplifiers) is properly calibrated.
Equation~\ref{eq:ss} is in the standard form of a linear dynamical
system $\dot{\vec V}=\bar A\vec V +\bar B\vec I^+$ where $\vec V$ is the column
vector with real and imaginary part of the voltages and $\vec I^+$ that of
the forward currents. The matrices $\bar A$ and $\bar B$ correspond to those in
Equation~\ref{eq:ss} and are given by
\begin{equation}\label{eq:AB}
  \bar A=\left(\begin{array}{cc} -\oo & -\Do \\ \Do& -\oo \end{array}\right)
  \qquad\mathrm{and}\qquad
  \bar B=\left(\begin{array}{cc} \ooe R & 0 \\ 0& \ooe R \end{array}\right) \ .
\end{equation}
For the simulations we will convert the continuous-time system from Equation~\ref{eq:ss}
to discrete time with time step $\Delta t$, which corresponds to the sampling time if the
system is implemented digitally. By replacing the derivatives of the voltages by
finite differences
\begin{equation}
  \frac{d\vec V}{dt}\to\frac{\vec V_{t+1}-\vec V_t}{\Delta t}
\end{equation}
where we label the time steps by $t$, Equation~\ref{eq:ss} becomes
\begin{equation}\label{eq:dsa}
  \vec V_{t+1} 
  =A \vec V_t  + B \vec I^+_t +\vec w_t
  \quad\mathrm{with}\quad A=\left(\begin{array}{cc} 1-\oo \dt & -\Do \dt\\ \Do \dt& 1-\oo \dt \end{array}\right)\ ,
\end{equation}
$B=\ooe \dt R\mathbf{1}$, and the process noise $\vec w_t$. We assume that the noise is
uncorrelated and has magnitude $\sigma_p$. It is thus characterized by its expectation
value $E\left\{\vec w_t\vec w_s^{\top}\right\}=\sigma_p^2\delta_{ts}\mathbf{1}$. We add measurement
noise $\vec w'_t$ by using
\begin{equation}
  \vec V'_t=\vec V_t + \vec w'_t
\end{equation}
in the system identification process.
We assume it is uncorrelated, has magnitude $\sigma_m$, and is characterized by
$E\left\{\vec w'_t\vec w_s^{\prime\top}\right\}=\sigma_m^2\delta_{ts}\mathbf{1}$.
\section{System identification}
\label{sec:sysid}
Now we turn to the task of extracting $\oo \dt$ and $\Do \dt$ from continuously
measured voltages $\vec V'_t$ and currents $\vec I^+_t$. In order to isolate the
sought parameters, we rewrite Equation~\ref{eq:dsa} in the form
\begin{equation}
  \vec V'_{t+1}=\left({\mathbf 1}+F\right)\vec V'_t + B\vec I^+_t
    \qquad\mathrm{with}\qquad
 F=\left(\begin{array}{rr} -\oo \dt & -\Do \dt\\ \Do \dt& -\oo \dt \end{array}\right)
\end{equation}
and $B=\ooe \dt R\mathbf{1}$. After reorganizing this equation to
\begin{equation}\label{eq:yy}
  \vec V'_{t+1}-\vec V'_t - B\vec I^+_t = F \vec V'_t 
\end{equation}
 we rewrite $F\vec V'_t$ on the right-hand side as
\begin{equation}\label{eq:yy2}
  F\vec V'_t = -\oo\dt \left(\begin{array}{c} V'_r \\ V'_i \end{array}\right)_t
  + \Do \dt \left(\begin{array}{r} -V'_i \\ V'_r \end{array}\right)_t
  = \left(\begin{array}{rr} -V'_r & -V'_i  \\ -V'_i & V'_r \end{array}\right)_t
  \left(\begin{array}{c} \oo \dt \\ \Do \dt \end{array}\right)\ .
\end{equation}
We now introduce the abbreviations
\begin{equation}\label{eq:defGy}
  G_t = \left(\begin{array}{rr} -V'_r & -V'_i  \\ -V'_i & V'_r \end{array}\right)_t
  \qquad\mathrm{and}\qquad
  \vec y_{t+1}= \vec V'_{t+1}-\vec V'_t- B\vec I^+_t
\end{equation}
and stack Equation~\ref{eq:yy} for consecutive times on top of each other.
In this way, we obtain a growing system of equations to determine
$\oo\dt$ and $\Do \dt$
\begin{equation}\label{eq:UT}
  \left(\begin{array}{c} \vec y_2 \\ \vec y_3 \\ \vdots \\ \vec y_{T+1}\end{array}\right)
  = U_T  \left(\begin{array}{c} \oo \dt \\ \Do \dt \end{array}\right)
  \quad\mathrm{with}\quad
  U_T=\left(\begin{array}{c} G_1 \\ G_2 \\ \vdots \\ G_T\end{array}\right)
\end{equation}
that we solve in the least-squares sense with the Moore-Penrose pseudo-inverse~\cite{PENROSE}
\begin{equation}\label{eq:ls}
  \vec q_T=\left(\begin{array}{c} \oo \dt \\ \Do \dt \end{array}\right)_T=
  \left( U_T^{\top} U_T\right)^{-1} U_T^{\top}
  \left(\begin{array}{c} \vec y_2 \\ \vec y_3 \\ \vdots \\ \vec y_{T+1}\end{array}\right) \ .
\end{equation}
Here we introduce the abbreviation $\vec q_T$ to denote the estimated parameters
at time step $T$.
\par
We can avoid lengthy evaluations by calculating Equation~\ref{eq:ls}
recursively. With the definition $P_T^{-1}=U^{\top}_TU_T$, its initial value
$P_0=p_0{\mathbf 1}$, and the definition of $U_T$ from Equation~\ref{eq:UT} we
express $P_{T+1}$ through $P_T$ in the following way
\begin{eqnarray}\label{eq:PTT}
  P_{T+1}^{-1}&=&U^{\top}_{T+1}U_{T+1}\nonumber\\
              &=&p_0{\mathbf 1}+G_1^{\top}G_1+G_2^{\top}G_2+\dots+G_T^{\top}G_T+G_{T+1}^{\top}G_{T+1}\\
              &=& P_T^{-1}+G_{T+1}^{\top}G_{T+1}\ . \nonumber
\end{eqnarray}
We note that for all time steps $t$
\begin{equation}\label{eq:ggone}
  G_t^{\top}G_t=(V_r^{\prime2}+V_i^{\prime2})_t{\mathbf 1} = \vec V_t^{\prime2} {\mathbf 1}
\end{equation}
is proportional to the unit matrix ${\mathbf 1}$. This renders the fit into two
orthogonal and independent parts; one for each of the fit parameters. To proceed, we
introduce the scalar quantity $p_T$ with $P_T=p_T{\mathbf 1}$ and find that it obeys
\begin{equation}
  p_{T+1}^{-1}=p_T^{-1}+\vec V_T^{\prime2}\ .
\end{equation}
Taking the reciprocal leads to
\begin{equation}\label{eq:upp}
  p_{T+1} = \left[\frac{1}{1+p_T\vec V_T^{\prime2}}\right]  p_T\ .
%  = \left[1 - \frac{p_T \vec V_T^{\prime2}}{1+p_T\vec V_T^{\prime2}}\right]p_T\ .
\end{equation}
Note that we need to initialize this recursion with a non-zero value and set
$p_0=1$ in the simulations. Despite being numerically unity, we carry $p_0$ through
all equations, because it carries the inverse units of $\vec V_T^2$.
\par
We now turn to finding $\vec q_{T+1}$ by writing Equation~\ref{eq:ls} for $T+1$
\begin{eqnarray}\label{eq:upq}
  \vec q_{T+1}&=& p_{T+1} \left(G_1^{\top}\vec y_2+G_2^{\top}\vec y_3+\dots
                  + G_T^{\top}\vec y_{T+1} +G_{T+1}^{\top}\vec y_{T+2}\right)\nonumber\\
              &=& \left[\frac{1}{1+p_T\vec V_T^{\prime2}}\right]
                  p_T\left(\sum_{t=1}^T G_t^{\top}\vec y_{t+1} +G_{T+1}^{\top}\vec y_{T+2}\right)\\
              &=&\left[\frac{1}{1+p_T\vec V_T^{\prime2}}\right]
                  \left( \vec q_T+p_TG_{T+1}^{\top}\vec y_{T+2}\right)\ . \nonumber
\end{eqnarray}
Equations~\ref{eq:upp} and~\ref{eq:upq} constitute the algorithm to continuously update
estimates for the two components of $\vec q$, the bandwidth $q(1)=\oo \dt$ and the
detuning $q(2)=\Do \dt$, as new voltage and current measurements--both enter in
$G_{T+1}$ and $\vec y_{T+2}$---become available. We refer to the MATLAB~\cite{MATLAB}
code on github~\cite{GITHUB} for the details of the implementation.
\par
In Equations~\ref{eq:upp} and~\ref{eq:upq} new information from measurements are used
to continuously improve the estimate of the fit parameters, but in situations where
they change, we have to introduce a way to forget old information. Therefore, 
in order to emphasize newly added information we follow~\cite{AW,OP} and introduce
a ``forgetting factor'' $\alpha=1-1/N_f$ where $N_f$ is the time horizon over
which old information is downgraded in the last equality of Equation~\ref{eq:PTT},
which now reads
\begin{equation}
P_{T+1}^{-1}= \alpha P_T^{-1}+G_{T+1}^{\top}G_{T+1}\ . \nonumber
\end{equation}
We see that we only have to replace $P_T$ by $P_T/\alpha$, or equivalently $p_T$ by
$p_T/\alpha$, in the derivation of Equations~\ref{eq:upp} and~\ref{eq:upq} and find
for the update of $p_T$
\begin{equation}\label{eq:uppt}
  p_{T+1} = \left[\frac{1}{\alpha+p_T\vec V_T^{\prime2}}\right]p_T
\end{equation}
and for the update of the estimated parameters $\vec q_T$
\begin{equation}\label{eq:upqt}
  \vec q_{T+1}=\left[\frac{1}{\alpha+p_T\vec V_T^{\prime2}}\right]
                 \left(\alpha \vec q_T+p_T\hat G_{T+1}^{\top}\vec y_{T+2}\right)
\end{equation}
that are capable of following time-dependent system parameters. These expressions
can be evaluated very efficiently. We find that the calculations in
Equation~\ref{eq:uppt} involve four multiplications and one inverse whereas the
calculations in Equation~\ref{eq:upqt} involve ten multiplications if we reuse
the expression in the square bracket. Thus, in total fourteen multiplication
and one, computationally more expensive, inverse are required. This is about
ten times the computational effort needed for a PI controller that typically
requires three multiplications. The processing delay of the system identification
algorithm should therefore be correspondingly longer. The details of the timing
depend of course on the hardware used to implement these algorithms. In particular,
on a field-programmable gate array, many operations can be done in parallel.
%..............................................
\begin{figure}[tb]
\begin{center}
\includegraphics[width=0.47\textwidth]{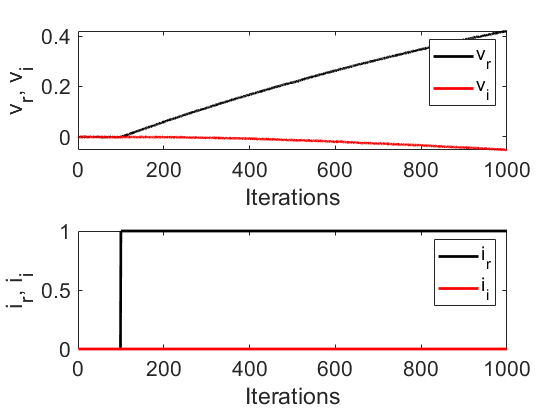}
\includegraphics[width=0.47\textwidth]{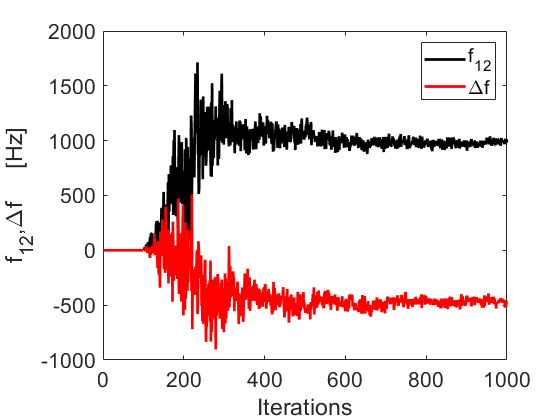}
\end{center}
\caption{\label{fig:Nf}Left: the normalized currents (bottom) and the voltages (top)
  after starting to fill the cavity for 1000 iterations ($100\,\mu$s). Right: the
  reconstructed fit parameters, the bandwidth $\ff$ (black) and the detuning $\Df$ (red).
  Note that the parameters are found despite the noise level ($\sigma_p=10^{-4}$
  and $\sigma_m=10^{-3}$ of peak voltage) used in the simulation.}
\end{figure}
%..............................................
%
\section{Simulations}
\label{sec:sim}
We base our simulations on parameters for the prototype spoke-cavity module~\cite{DUCHESNE}
for the ESS~\cite{ESS}, which operates at 352\,MHz, has an external $Q_E$~\cite{HANLI21} in the
range $1.75\times 10^5$ to $2.85\times10^5$. One of the measured cavities exhibited a loaded-$Q$
of $Q_L=1.8\times 10^5$~\cite{HANLI} while it was operating at a high gradient. The resulting
bandwidth is $f_{12}=\oo/2\pi \approx 1000\,$Hz. The cavity showed Lorentz-force detuning on
the order of a few hundred Hz~\cite{HANLI,HANLI19,ROCIO}; for our simulations we typically use
$\Delta f =\Do/2\pi=500\,$Hz.
Moreover, we use a process noise level of $\sigma_p=10^{-4} \times V_{max}$ and a measurement
noise level of $\sigma_m=10^{-3} \times V_{max}$, where $V_{max}$ is the peak
voltage inside the cavity. We report the voltages and currents normalized to the values without
detuning and denote them by $v_r,v_i$ and $i_r,i_i$ respectively. The peak voltage and current
in those conditions then becomes unity. Furthermore, we assume that the data-acquisition system
operates at a rate of 10\,Msamples/s, resulting in $\Delta t=100\,$ns. We found that the
forgetting horizon $N_f$ scales with the relative noise levels $\sigma_p$ and $\sigma_m$. 
We use $N_f=100$, unless explicitely specified, because it gave good results.
\par
The left-hand side in Figure~\ref{fig:Nf} shows the normalized currents and voltages over
the first 1000 iterations ($100\,\mu$s), where the currents are turned on after 100 iterations.
We observe that the real part of the current (black line) assumes its new value at that point
whereas the imaginary part (red line) stays zero. The voltages, shown on the upper panel
slowly starts rising as the cavity is filled. Even the imaginary part of the voltage
deviates from zero, owing to the finite value of the detuning. The right-hand side of
Figure~\ref{fig:Nf} shows the fit parameters $\ff=\oo/2\pi$ and $\Df=\Do/2\pi$ over the
same 1000 iterations. We observe that during the first few hundred iterations the estimated
fit parameters are very noisy, but settle on their correct value after this initial period.
After about iteration 600 they meander quite closely around their ``true'' values.
\par
%..............................................
\begin{figure}[tb]
  \begin{center}
    \includegraphics[width=0.7\textwidth]{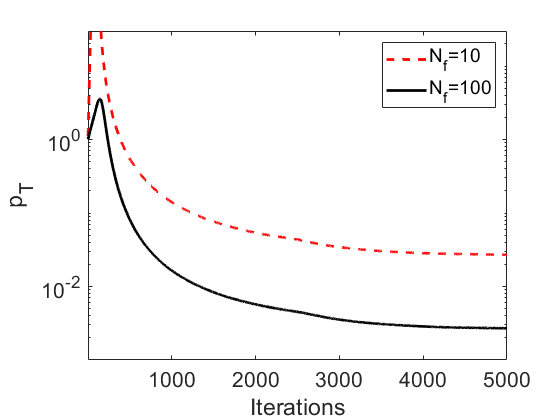}
  \end{center}
  \caption{\label{fig:pt}The variable $p_T$ as a function of the iterations for $N_f=100$
    (solid) and $N_f=10$ (dashes).}
\end{figure}
%..............................................
We can understand this behavior by noting that $p_T$ is proportional to the diagonal element
of the empirical covariance matrix $P_T=\left(U_T^{\top}U_T\right)^{-1}$ of the least-squares fit
in Equation~\ref{eq:ls}. Therefore the square root of $p_T$ is proportional to the error bars
of the fit parameter. Figure~\ref{fig:pt} shows $p_T$ for a simulation with $N_f=100$ (black solid)
and $N_f=10$ (red dashes) for 5000 iterations. We observe that both curves initially increase
during the period that the fit is noisy but then approach a constant value that determines
the achievable error bars of the fit parameters. This value can be derived from Equation~\ref{eq:uppt}
by setting $p_{T+1}=p_T=p_{\infty}$ and solving for $p_{\infty}=1/N_f\vec V^{\prime2}_{\infty}$. Here $\vec V'_{\infty}$
is the voltage inside the cavity. For the error bars of both components of $\vec q$ we thus find
$\sigma_m/\sqrt{N_f\vec V^{\prime 2}_{\infty})}$, a value that corresponds to the rms deviations of the fit
parameters, shown, for example on the second half in Figure~\ref{fig:Nf}. Furthermore by construction,
the off-diagonal elements of the matrix $P_T=p_T\mathbf{1}$ are zero, which indicates that the
fit of the bandwidth and the detuning are orthogonal and that makes the algorithm very robust.
Moreover, we found that instead of operating open-loop, using a PI-controller to control the
cavity voltage does not significantly alter the performance of the system identification process.
\par
%..............................................
\begin{figure}[tb]
  \begin{center}
    \includegraphics[width=0.47\textwidth]{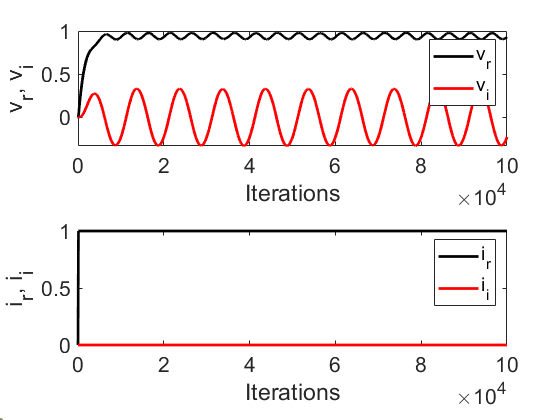}
    \includegraphics[width=0.47\textwidth]{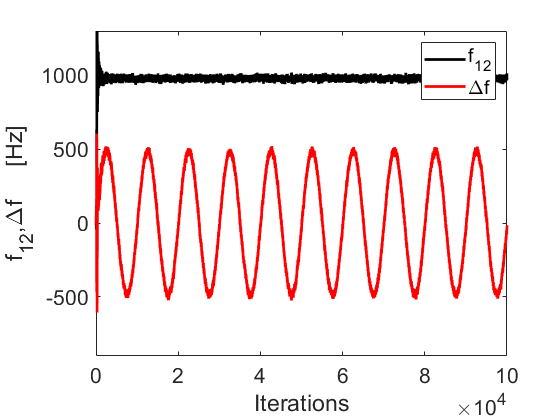}
  \end{center}
\caption{\label{fig:osc}The currents and voltages (left) and the fit parameters
  (right) for $10^5$ iterations (10 ms) while the detuning $\Df$ oscillates with an amplitude of 500\,Hz and
  with a mechanical-mode frequency of 1\,kHz. The oscillations are clearly visible
  on both phases of the voltage and the correctly reconstructed fit parameters.}
\end{figure}
%..............................................
We now explore the algorithm's ability to identify parameter changes during steady state
operation. The left-hand side in Figure~\ref{fig:osc} illustrates the effect of microphonics
on the the currents and voltages. We simulate this by an oscillation of $\Df$ with
amplitude of $\ff/2$ and frequency 1\,kHz. Especially $v_i$ reveals this oscillation,
though also $v_r$ oscillates. The right-hand side of Figure~\ref{fig:osc} shows how the
algorithm correctly identifies $\ff$ and both the amplitude and oscillation frequency of $\Df$.
\par
%..............................................
\begin{figure}[tb]
  \begin{center}
    \includegraphics[width=0.7\textwidth]{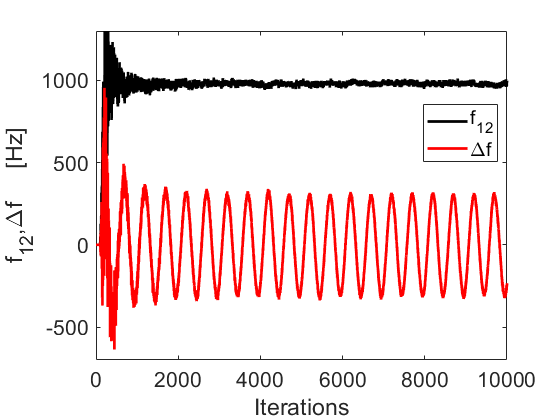}
  \end{center}
  \caption{\label{fig:otherp}The reconstructed fit parameters for a 20\,kHz mechanical
    oscillation of the detuning $\Df$. The oscillations are still seen, but the amplitude
    is significantly reduced. This can be partially alleviated by decreasing $N_f$,
    albeit at the expense of an increased noise level.}
\end{figure}
%..............................................
Increasing the oscillation frequency to 20\,kHz results in Figure~\ref{fig:otherp} where
we have reduced the duration of the simulation to $10^4$ iterations in order to improve
the visibility of oscillations on the plot. We see that the oscillations are still resolved,
albeit at a lower amplitude, which is a consequence of the forgetting horizon $N_f=100$.
It implicitly introduces averaging over $N_f$ iterations and thus behaves like a low-pass
filter with a time constant of $N_f\Delta t=10\,\mu$s or a cutoff frequency on the order
of 100\,kHz that already causes some attenuation of the 20\,kHz oscillation.
\par
%..............................................
\begin{figure}[tb]
  \begin{center}
    \includegraphics[width=0.47\textwidth]{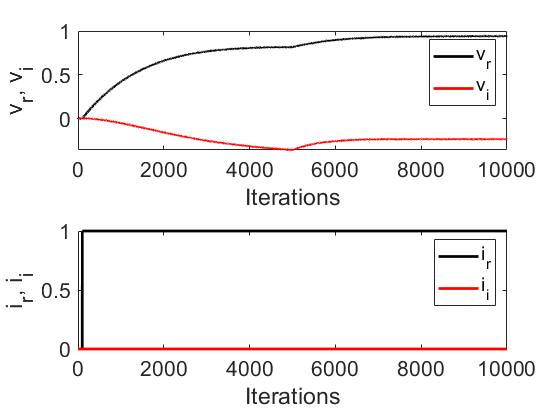}
    \includegraphics[width=0.47\textwidth]{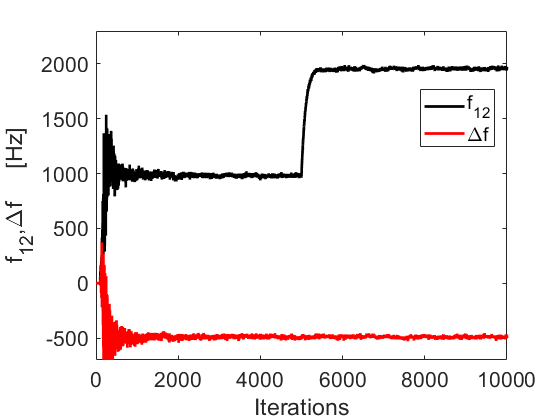}
    \includegraphics[width=0.47\textwidth]{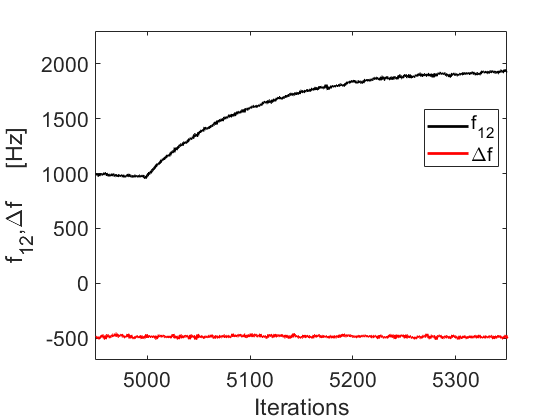}
    \includegraphics[width=0.47\textwidth]{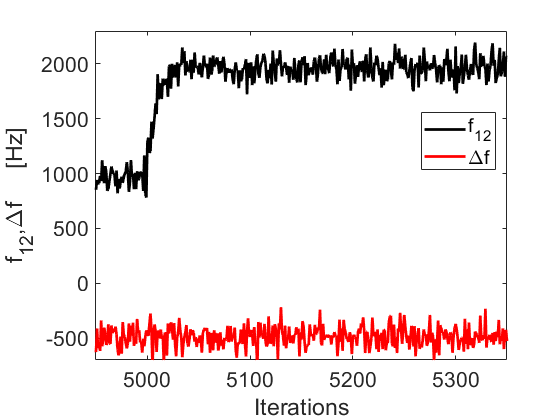}
  \end{center}
  \caption{\label{fig:step}The currents and voltages (top left) and fit parameters (top right)
    for $10^4$ iterations (1\,ms) as the bandwidth $\ff$ is doubled at iteration 5000. The
    bottom row shows an enlarged view of fit parameters around the time of the change. On
    the left, we use $N_f=100$ and on the right we use $N_f=10$.}
\end{figure}
% ..............................................
In Figure~\ref{fig:step} we explore a rapid increase of the bandwidth, for example, due to
a quench. In the simulation, we simply double the value of $\oo$ after 5000 iterations.
The plots in the top-left of Figure~\ref{fig:step} show the currents and voltages and
on the top-right the fit parameters. We find that the fitted bandwidth (black) is indeed
doubled and that the reconstruction of the detuning is unaffected. The plot on the bottom
left shows an enlarged view of the fit parameters around the time of the step. It shows
that the doubled value is approached within about $2\times N_f=200$ iterations. If we run
the same simulation with a ten times reduced value of $N_f=10$, we obtain the plot on
the bottom right. We find that the changed value is approached within a few tens of
iterations, albeit at the expense of an increased noise level, which is consistent with
the discussion regarding Figure~\ref{fig:pt}. Balancing the noise level and the response
is just a matter of adjusting the value of $N_f$, the topic of the following section.
\section{Signal to Noise}
\label{sec:SN}
In Section~\ref{sec:sim} we already found that the asymptotic noise level $N$ for constant
parameters is given by
\begin{equation}
N=\frac{1}{\sqrt{N_f}}\frac{\sigma_m}{V'_{\infty}}\ ,
\end{equation}
where we denote the magnitude of $\vec V'_{\infty}$ by $V'_{\infty}$. We now consider a
situation where the system has reached a quasi-stationary state and that perturbations
of the $\oo$ and $\Do$ are so small that they affect $V'_{\infty}$ very little. We can
therefore also use it to write $p_{\infty}=1/N_fV^{\prime 2}_{\infty}$ despite temporally
varying $\oo$ and $\Do$. Replacing $p_T$ by $p_{\infty}$ in Equation~\ref{eq:upqt} then
leads to
\begin{equation}
  \vec q_{T+1} = \alpha\vec q_T + \frac{1}{N_fV^{\prime 2}_{\infty}}G^{\top}_{T+1}\vec y_{T+2}\ .
\end{equation}
Using Equation~\ref{eq:yy} and~\ref{eq:yy2} we rewrite $\vec y_{T+2}$ as
\begin{equation}
  \vec y_{T+2} = G_{T+1} \left(\begin{array}{c} \oo \dt \\ \Do \dt \end{array}\right)_{hw}
\end{equation}
where the vector on the right-hand side with the subscript $hw$ are the ``true''
values of the hardware. Combining these equations, utilizing Equation~\ref{eq:ggone},
and replacing $V'_T$ by $V'_{\infty}$ we arrive at
\begin{equation}
 \vec q_{T+1} = \alpha\vec q_T + \frac{1}{N_f}\left(\begin{array}{c} \oo \dt \\ \Do \dt \end{array}\right)_{hw}\ .
\end{equation}
In the next step we use $\alpha=1-1/N_f$ and reshuffle terms to obtain
\begin{equation}
  \frac{\vec q_{T+1}-\vec q_{T}}{\Delta t} = -\frac{1}{N_f\Delta t} \vec q_T - \frac{1}{N_f\Delta t}
  \left(\begin{array}{c} \oo \dt \\ \Do \dt \end{array}\right)_{hw}\ .
\end{equation}
Introducing $\tau_f=N_f\Delta t$, replacing the finite difference by a differential,
and Laplace-transforming the resulting equation we find
\begin{equation}
  \left( s+\frac{1}{\tau_f}\right) \tilde{\vec q} = \frac{1}{\tau_f}
  \left(\begin{array}{c} \tilde\oo \dt \\ \Delta\tilde\omega \dt \end{array}\right)_{hw}
\end{equation}
where $s$ is the Laplace variable and we denote the Laplace transform of a variable by a tilde.
We obtain the the time dependence by replacing $s=i\omega=2\pi i f$
\begin{equation}
 \tilde{\vec q} = \frac{1}{1+i\omega\tau_f}\left(\begin{array}{c} \tilde\oo \dt \\ \Delta\tilde\omega \dt \end{array}\right)_{hw}
\end{equation}
and find that the reconstructed system parameters $\tilde{\vec q}$ are given by
the hardware parameters passed through a low-pass filter with time constant $\tau_f$.
\par
Of particular interest is the absolute value of the amplitude of the detuning $\Delta\tilde\omega$
at frequency $\omega$, which is given by
\begin{equation}
S=\Delta\tilde\omega = \frac{\Delta\tilde\omega_{hw}}{\sqrt{1+(\omega\tau_f)^2}}\ .
\end{equation}
This constitutes the signal we strive to measure. For the signal-to-noise ratio $S/N$
we then find
\begin{equation}\label{eq:SN}
  S/N= \frac{\Delta\tilde\omega_{hw}}{\sqrt{1+\left(2\pi N_f f \Delta t\right)^2}}
  \frac{\sqrt{N_f}}{(\sigma_m/V'_{\infty})}\ ,
\end{equation}
where all parameters are explicitely written out in order to explore the trade-off
among them. Apparently it depends on the magnitude (amplitude) of the detuning
$\Delta\tilde\omega_{hw}$ and the relative accuracy of the voltage measurement
$\sigma_m/V'_{\infty}$, but also on the attenuation of an oscillation due to the
forgetting time horizon $N_f$. As long as $S/N$ is sufficiently large, say $5$
or so, the oscillation is discernible.
\section{Conclusions}
We worked out an algorithm to determine the cavity bandwidth $\ff$ and the detuning
$\Df$ by correlating signal from a directional coupler before the cavity and the
voltages inside the cavity. The calculations are very efficient and given by
Equations~\ref{eq:upp} and~\ref{eq:upq} for static parameters and by Equations~\ref{eq:uppt}
and~\ref{eq:upqt} for time-varying parameters. These recursion equations are very
compact and require only moderate resources, for example, on a field-programmable
gate array.
\par
Despite the absence of low-pass filtering, the RLS algorithm is resilient to noise
of the measured voltages, because the forgetting horizon implicitly introduces a
low-pass filter whose time constant is $\tau_f=N_f\dt$. We can taylor
the performance by selecting a large value of $N_f$, which reduces the noise of the
reconstructed parameters, whereas smaller values of $N_f$ make the algorithm more
responsive to parameter changes on faster times scales. The trade-off between
achievable frequency resolution, $N_f$, and measurement noise $\sigma_m$ can be
explored with the help of Equation~\ref{eq:SN}.
\subsection*{Acknowledgments}
Discussions with Tor Lofnes, Uppsala University are gratefully acknowledged.
% 
%%%%%%%%%%%%%%%%%%%%%%%%%%%%%%%%%%%%%%%%%%%%%%%%%%%%%%%%%%%%%%%%%%%%%%%
%
\bibliographystyle{plain}

\begin{thebibliography}{M}
  %
\bibitem{SNS}
  S. Henderson et al., {\em The Spallation Neutron Source accelerator system design,}
  Nucl. Instrum. Methods A763 (2014) 610.
\bibitem{ESS}
  A. Jansson et al., {\em The status of the ESS project,} Proceedings of IPAC 2022,
  Bangkok, 792.
\bibitem{CEBAF}
  B. Norum, J. McCarthy, R. York, {\em CEBAF - a high-energy, high duty factor
    electron accelerator for nuclear physics,} Nucl. Instrum. Methods B10/11 (1985) 337.
\bibitem{XFEL}
  W. Decking et al., {\em A MHz-repetition-rate hard X-ray free-electron laser driven
    by a superconducting linear accelerator,} Nature Photonics 14 (2020) 391. 
\bibitem{CBETA}
  A. Bartnik, N. Banerjee, D. Burke, J. Crittenden, K. Deitrick, J. Dobbins, et al.,
  {\em CBETA: First Multipass Superconducting Linear Accelerator with
    Energy Recovery,} Phys. Rev. Lett. 125, 044803, July 2020.
\bibitem{FRIB}
  P. Ostroumov et al., {\em Beam commissioning in the first superconducting segment
    of the Facility for Rare Isotope Beams,} Phys. Rev. Accel. Beams 22, 080101.
\bibitem{SPIRAL2}
  H. Goutte, A. Navin, {\em Microscopes for the Physics at the Femtoscale: GANIL-SPIRAL2,}
  Nucl.Phys.News 31 (2021) 1, 5.
\bibitem{HELIAC}
  W. Barth et al., {\em Advanced basic layout of the HElmholtz LInear ACcelerator for
    cw heavy ion beams at GSI,} Proceedings of IPAC 2023, Venice,
  https://doi.org/10.18429/JACoW-IPAC2023-TUPA186
\bibitem{JLABFEL}
  C. Behre et al., {\em First lasing of the IR upgrade FEL at Jefferson lab,}
  Nucl. Instrum. Methods A528 (2004) 
\bibitem{LFD}
  B. Aune et al., {\em Superconducting TESLA cavities,} Phys. Rev. ST Accel. Beams 3 (2000) 092001.
\bibitem{ACE3}
  O. Kononenko, C. Adolphsen, Z. Li, C. Ng, C. Rivetta, {\em 3D multiphysics modeling of superconducting
    cavities with a massively parallel simulation suite,} Phys. Rev. Accel. Beams 20 (2017) 102001.
\bibitem{ANA1}
  G. Davis et al., {\em Microphonics Testing of the CEBAF Upgrade 7-Cell Cavity,} Proceedings of PAC 2001, 1152.
\bibitem{NEUMANN}
  A. Neumann, W. Anders, O. Kugeler, J. Knobloch, {\em Analysis and active compensation of microphonics
    in continuous wave narrow-bandwidth superconducting cavities,} Phys. Rev. ST Accel. Beams 13 (2010) 082001. 
\bibitem{TUNER}
  M. Liepe, {\em Superconducting Multicell Cavities for Linear Colliders,} Dissertation, Universit\"at Hamburg, 2001.
\bibitem{ANA2}
  T. Powers, {\em Theory an practice of Cavity RF test systems,} Proceedings of the 12th
  International Workshop on RF Superconductivity, Cornell University, Ithaca, New York, USA (2005) 30.
\bibitem{SCHILCHER}
  T. Schilcher, {\em Vector sum control of pulsed accelerating fields in lorentz
    force detuned superconducting cavities,} Dissertation, Universit\"at Hamburg, 1998.
\bibitem{PLAWSKI}
  T. Plawski et al., {\em Digital Cavity Resonance Monitor-Alternative way to Measure Cavity
    Microphonics,} Proceedings of the 12th International Workshop on RF Superconductivity,
  Cornell University, Ithaca, New York, USA (2005) 616.
\bibitem{RYBA}
  R. Rybaniec et al., {\em Real-time estimation of superconducting cavities parameters,}
  Proceedings of EPAC 2014 in Dresden, p. 2456.
\bibitem{CZARSKI}
  T. Czarski, {\em Superconducting cavity control based on system model identification,}
  Meas. Sci. Technol. {\bf 18} (2007) 2328.
\bibitem{BELL}
  A. Bellani et al., {\em Online detuning computation and quench detection for superconducting
    resonators,} IEEE Transactions on Nuclear Science 68 (2021) 385.
\bibitem{ECHE}
  P. Echevarria, B. Arruabarrena, A. Ushakov, J. Jugo, A. Neumann, {\em Simulation of quench
    detection algorithms for Helmholtz Zentrum Berlin SRF cavities,} Proceedings of the
  tenth International Particle Accelerator Conference in Melbourne (2019) 2834.
\bibitem{AW}
  K. \AA str\"om, B. Wittenmark, {\em Adaptive Control, 2nd edition,} Dover Publications,
  Mineola, 2008; especially Section 2.2.
\bibitem{ZZ1}
  I. Ziemann, V. Ziemann, {\em Noninvasively improving the orbit-response matrix while continuously
    correcting the orbit,} Physical Review Accelerators and Beams 24 (2021) 072804.
\bibitem{LAIWEI}
  T. Lai, C. Wei, {\em Least squares estimates in stochastic regression models with
    applications to identification and control of dynamic systems,} The Annals of
  Statistics 10 (1982) 143.
% \bibitem{VZAPB}
%   V. Ziemann, Hands-On Accelerator Physics Using MATLAB, CRC Press, Boca Raton, 2019;
%   especially Section 6.6.3.
% \bibitem{FYF}
%   V. Ziemann, {\em Physics and Finance,} Springer, Heidelberg, 2021.
% \bibitem{KIRK}
%   D. Kirk, {\em Optimal control theory,} Dover, New York, 2004.
\bibitem{NR}
  W. Press et al.,{\em Numerical Recipes, 2nd ed.}, Cambridge University Press, Cambridge, 1992.
\bibitem{PENROSE}
  R. Penrose, {\em A generalized inverse for matrices,} Mathematical Proceedings of the Cambridge
  Philosophical Society {\bf 51} (1955) 406 (\url{https://doi.org/10.1017/S0305004100030401}).
\bibitem{MATLAB}
  Mathworks web site at \url{www.mathworks.com}
\bibitem{GITHUB}
  Github repository for the software accompanying this report: \url{https://github.com/volkziem/SysidRFcavity}.
\bibitem{OP}
  V. Ziemann, {\em Operational improvements for an algorithm to noninvasively measure the orbit
    response matrix in storage rings,} arXiv:2303.11216, March 2023.
\bibitem{DUCHESNE}
  P. Duchesne, S. Bousson, S. Brault, P. Duthil, G. Olry, D. Reynet, S. Molloy,
  {\em Design of the 352 MHz, beta 0.50, double-spoke cavity for ESS,} Proceedings of SRF2013
  (2013) 1212.
\bibitem{HANLI21}
  H. Li, A. Miyazaki, M. Zhovner, L. Hermansson, R. Santiago Kern, K. Fransson, K. Gajewski, R. Ruber,
  {\em Progress and preliminary statistics for the ESS series spoke cryomodule test,} Proceedings
  of SRF2021 (2021) 512.
\bibitem{HANLI}
  H. Li, A. Miyazaki, R. Santiago Kern, L. Hermansson, T. Lofnes, K. Gajewski, K. Fransson,
  R. Wedberg, R. Ruber, {\em RF performance of the spoke prototype cryomodule for ESS,}
  FREIA Report 2019/08, 2019.
\bibitem{HANLI19}
  H. Li, M. Jobs, R. Santiago Kern, V.A. Goryashko, L. Hermansson, A. Bhattacharyya, T. Lofnes,
  K. Gajewski, K. Fransson, R. Ruber, {\em Characterization of a $\beta=0.5$ double spoke cavity
    with a fixed power coupler,} Nuclear Inst. and Methods in Physics Research, A 927 (2019) 63.
\bibitem{ROCIO}
  R. Santiago Kern, C. Svanberg, K. Fransson, K. Gajewski, L. Hermansson, H. Li, T. Lofnes,
  M. Olveg\aa rd, I. Profatilova, M. Zhovner, A. Miyazaki, R. Ruber,
  {\em Completion of Testing Series Double-spoke Cavity Cryomodules for ESS,}
  presented at the 21st International Conference on Radio-Frequency Superconductivity (SRF 2023);
  see also \url{https://arxiv.org/abs/2306.11333}.
%\bibitem{CLOP}
%  V. Ziemann, {\em Real-time system identification of superconducting cavities with
%    a recursive least-squares algorithm: closed-loop operation}, arXiv:2307.07386, July 2023.
%   available from \url{https://doi.org/10.48550/arXiv.2307.07386}.
% \bibitem{ABRASTE}
%   M. Abramowitz, I. Stegun, {\em Handbook of Mathematical Functions,} Dover, New York, 1972.
% %
\end{thebibliography}

%
%%%%%%%%%%%%%%%%%%%%%%%%%%%%%%%%%%%%%%%%%%%%%%%%%%%%%%%%%%%%%%
%\appendix
%
% 
\end{document}